\documentclass[aps,pra,twocolumn,groupedaddress,showpacs,floatfix]{revtex4}
\usepackage{graphicx}
\usepackage{amsmath}

\begin{document}

\title{Anomalous Minimum in the Shear Viscosity of a Fermi Gas}

\author{E. Elliott$^{1,2}$, J. A. Joseph$^1$, and J. E. Thomas$^1$}

\affiliation{$^{1}$Department of  Physics, North Carolina State University, Raleigh, NC 27695, USA}
\affiliation{$^{2}$Department of Physics, Duke University, Durham, NC 27708, USA}


\date{\today}

\begin{abstract}
We measure the static shear viscosity $\eta$ in a two-component Fermi gas near a broad collisional (Feshbach) resonance, as a function of interaction strength and energy.  We find that $\eta$  has both a quadratic and a linear dependence on the interaction strength $1/({k_{FI}a})$, where $a$ is the s-wave scattering length and $k_{FI}$ is the Fermi wave vector for an ideal gas at the trap center.  For energies above the superfluid transition, the minimum in $\eta$ as a function of interaction strength is significantly shifted toward the BEC side of resonance, to $1/(k_{FI}a)\simeq 0.25$.
\end{abstract}

\maketitle

Ultra-cold Fermi gases provide a unique model for studying the properties of strongly interacting quantum fluids~\cite{OHaraScience,RMP2008,ZwierleinFermiReview,ZwergerReview,NJPReview}.  Utilizing a collisional (Feshbach) resonance, a bias magnetic field readily tunes interactions between spin-up and spin-down atoms  from  non-interacting to strongly repulsive or strongly attractive~\cite{BartensteinFeshbach,JochimFeshbach}. Several ground-breaking  measurements have focused on the  equilibrium thermodynamic properties~\cite{LuoEntropy,ThermoLuo,ThermoUeda,ThermoSalomon,KuThermo}. However, systematic study of interaction-dependent hydrodynamic transport coefficients poses new challenges. Measurement of the shear viscosity is of particular interest in recent predictions~\cite{BruunViscousNormalDamping, TaylorViscosity,LevinViscosity,ZwergerViscosity,BulgacShear,BulgacTempShearUnitary} and in the context of a ``perfect" fluid conjecture~\cite{Kovtun}, derived using holographic duality methods~\cite{NJPReview}. The conjecture  states that for a broad class of (conformal) strongly interacting quantum fields, the ratio of the shear viscosity $\eta$ to the entropy density $s$ has a universal minimum, $\eta/s\geq\hbar/(4\pi k_B)$~\cite{Kovtun}.  Recent measurements of the shear viscosity for a resonantly interacting Fermi gas~\cite{CaoViscosity,CaoNJP} yield a minimum $\eta/s$ ratio just $4.5$ times the lower bound, comparable to that of a quark-gluon plasma~\cite{NJPReview}. Whether the shear viscosity of a Fermi gas (or the $\eta/s$ ratio) is minimized at resonance or at a finite scattering length is an open question.

In this Letter, we describe a measurement of the shear viscosity $\eta$ in an expanding Fermi gas as a function of the interaction strength and energy near a broad Feshbach resonance~\cite{BartensteinFeshbach,JochimFeshbach}. The shear viscosity is determined with high sensitivity by releasing the cloud from a cigar-shaped  optical trap  with an elliptical (1:2.7) transverse profile and measuring the transverse aspect ratio as a function of time after release~\cite{ElliottScaleInv}. The interaction strength is characterized by the dimensionless parameter $1/(k_{FI} a)$, where $a$ is the s-wave scattering length and $k_{FI}$ is the Fermi wavevector of an ideal gas at the trap center. First we determine the shear viscosity at resonance, where $1/(k_{FI} a)=0$, and then we determine the correction to the shear viscosity as a function of $1/(k_{FI} a)$. In kinetic theory, the correction term is expected to scale as  $1/(k_{FI}a)^2$~\cite{BruunViscosity2}. However, for a given energy, we find there is an additional linear dependence on $1/(k_{FI}a)$ that results in a shift of the minimum viscosity.

For the experiments, we employ a Fermi gas of $^6$Li atoms in a 50-50 mixture of the two lowest hyperfine states, which is confined in a cigar-shaped optical trap with aspect ratios x:\thinspace y:\thinspace z = 1:\thinspace 2.7:\thinspace 33. The cloud is tuned near a broad Feshbach resonance and cooled by evaporation~\cite{OHaraScience}.
After evaporative cooling, the interaction strength is adjusted by tuning the bias magnetic field. Then the optical trap is extinguished  and the cloud radii are measured as a function of time after release in all three dimensions, using two simultaneous probe pulses interacting with different spin states to obtain independent absorption images on two CCD cameras~\cite{ElliottScaleInv}.

We define a general, scattering-length-independent, energy scale $\widetilde{E}$ for the trapped cloud  by
\begin{equation}
\widetilde{E}\equiv\frac{3}{N}\int d^3{\mathbf{r}}\,p_{\,0}=\langle{\mathbf{r}}\cdot\nabla U\rangle_0.
\label{eq:tildeE}
\end{equation}
Here, $p_{\,0}$ is the equilibrium pressure and $\widetilde{E}$  is then three times the grand potential per particle~\cite{ResonantE}. The form on the right follows from force balance, $\nabla p_0+n_0\,\nabla U=0$, with $n_0$ the equilibrium density.   $\tilde{E}$ is given by the trap average \mbox{$\langle{\mathbf{r}}\cdot\nabla U\rangle_0\equiv\frac{1}{N}\int d^3{\mathbf{r}}\,n_0({\mathbf{r}})\,{\mathbf{r}}\cdot\nabla U$}. For the measured trap parameters, given below, $\tilde{E}$ is then determined  by  the measured spatial profile of the trapped cloud~\cite{SupportOnline}. Hence, by fixing $\widetilde{E}$, we fix the average density for our measurements of viscosity at different interaction strengths.

The total trapping potential $U=U_{opt}+U_{mag}$ contains an optical part $U_{opt}$ and a magnetic part $U_{mag}$, arising from curvature in the bias magnetic field. For the optical potential, we find: $\omega_{x}=2\pi\times 2210(4)$ Hz, $\omega_{y}=2\pi\times 830(2)$ Hz, and $\omega_{zopt}=2\pi\times 60.6(0.4)$ Hz. The additional magnetic potential $U_{mag}=\frac{1}{2}m\,\omega_{mag}^2(y^2+z^2-2x^2)$, where $m$ is the $^6$Li mass and $\omega_{mag}=2\pi\times 21.5(0.25)\sqrt{B/834}$ Hz is the oscillation frequency of the cloud along the $y$-axis, which is measured at 834 G with $U_{opt}=0$. For later use, we define the ideal gas Fermi energy $E_F\equiv (3N)^{1/3}\hbar\bar{\omega}$, and the corresponding wavevector $k_{FI}=(2mE_F/\hbar^2)^{1/2}$, where $N$ is the total number of atoms, which is typically $2\times 10^5$ and $\bar{\omega}=(\omega_{x}\omega_{y}\omega_z)^{1/3}$ with $\omega_z=(\omega^2_{zopt}+\omega^2_{mag})^{1/2}$.

After release from the cigar-shaped trap, the transverse aspect ratio of the cloud exhibits elliptic flow in the x-y plane, indicating hydrodynamic expansion~\cite{ElliottScaleInv}. The shear viscosity pressure tensor slows the flow in the initially narrow, rapidly expanding, x-direction and transfers energy to the more slowly expanding y-direction. For a fixed time after release, the transverse aspect ratio $\sigma_x/\sigma_y$ then decreases with increasing shear viscosity. In contrast to elliptic flow measurements employing the axial z-direction, which expands slowly, the relatively high frequencies $\omega_x$ and $\omega_y$ assure that $\sigma_x/\sigma_y$ saturates on a rapid time scale, where  the expanded cloud images still have high signal to background ratio, and  reduces  sensitivity to the magnetic potential.

The shear viscosity is given in natural units of $\hbar \,n$ by $\eta\equiv\alpha_S\,\hbar\,n$, where $n$ is the density and $\alpha_S$ is a dimensionless shear viscosity coefficient~\cite{CaoViscosity}. The transverse aspect ratio data is fit using a hydrodynamic model, described below, to determine the {\it cloud-averaged} shear viscosity coefficient $\langle\alpha_S\rangle$, where
\begin{equation}
\langle\alpha_S\rangle\equiv\frac{1}{N\hbar}\int d^3{\mathbf{r}}\,\eta=\frac{1}{N}\int d^3{\mathbf{r}}\, n\,\alpha_S.
\label{eq:trapavcoeff}
\end{equation}
At finite scattering length, $\langle\alpha_S\rangle$ is generally time dependent, as discussed further below.

For our experiments below resonance,  at low temperatures, a BEC would exist, and a two-fluid description would be required. To consistently compare our measurements of shear viscosity throughout the resonance region, we therefore work in the normal fluid regime,  avoiding complications arising from two-fluid behavior that is not observed for the conditions of our experiments. Further, we estimate that the ratio of the collisional mean free path  to the cloud size (the Knudsen number) is  small for both the molecular and atomic components~\cite{PetrovDimerDimerScattLength,PetrovAtomDimerScattLength,SupportOnline}.
Hence, we determine the trap-averaged shear viscosity coefficients by fitting a hydrodynamic theory for a single component fluid.

For a single component fluid~\cite{StringariBulk,SupportOnline}, the velocity field $\mathbf{v}(\mathbf{r},t)$ obeys the Navier-Stokes equation~\cite{LandauFluids}, which includes the scalar pressure $p$, the trap potential $U$, and generally the shear and bulk viscosities~\cite{SupportOnline}. With current conservation, we obtain exact evolution equations~\cite{NeglectBulk} for the mean square cloud sizes $\langle x_i^2\rangle$, $i=x,y,z$,
\begin{equation}
\frac{d^2}{dt^2}\frac{m\langle x_i^2\rangle}{2}=\frac{1}{N}\!\int\!\!d^3{\mathbf{r}}\thinspace p+m\langle v_i^2\rangle-\langle x_i\partial_i U\rangle -\hbar\langle\alpha_S\sigma_{ii}\rangle,
\label{eq:xsqddot2}
\end{equation}
where $\langle ...\rangle$ denotes an average over the cloud density, as in Eq.~\ref{eq:trapavcoeff},   and $\sigma_{ij}=\partial v_i/\partial x_j+\partial v_j/\partial x_i-2\delta_{ij}\nabla\cdot\mathbf{v}/3$.

We see that the pressure $p$ in Eq.~\ref{eq:xsqddot2} arises only in a volume integral, which we determine using energy conservation. After release of the cloud, when $U$ is temporally constant, we have~\cite{SupportOnline},
\begin{equation}
\frac{d}{d t}\int d^3{\mathbf{r}}\,{\cal E}+\int d^3{\mathbf{r}}\,(\nabla\cdot{\mathbf{v}})\,p\, =\dot{Q},
\label{eq:3.1e}
\end{equation}
where ${\cal E}$ is the energy density and $\dot{Q}$ is the total heating rate arising from the viscous forces~\cite{SupportOnline}. Eq.~\ref{eq:3.1e} is used to find $\frac{1}{N}\!\int\!\! d^3{\mathbf{r}}\thinspace p$ by eliminating ${\cal E}$, using $p=\frac{2}{3}{\cal E}+\Delta p$, where $\Delta p$ is the predicted conformal symmetry breaking pressure, which vanishes at resonance~\cite{HoUniversalThermo,ElliottScaleInv}. Note that the measured $\widetilde{E}$ in Eq.~\ref{eq:tildeE} determines the initial condition, $\frac{1}{N}\!\int\!\! d^3{\mathbf{r}}\thinspace p_0$.

To solve Eqs.~\ref{eq:xsqddot2}~and~\ref{eq:3.1e}, we use a scaling approximation, which has been shown to be very accurate using numerical viscous hydrodynamics~\cite{SchaeferDissipativeScaling}. Then,  $\langle x_i^2\rangle=\langle x_i^2\rangle_0\,b_i^2(t)$ and $\langle v_i^2\rangle=\langle x_i^2\rangle_0\,\dot{b_i}^2(t)$, where $b_x,b_y,b_z$ are the expansion scale factors and $\langle x_i^2\rangle_0$ is the measured mean square size just after release.  In the scaling approximation, $v_i=x_i\dot{b}_i/b_i$ and $\nabla\cdot{\mathbf{v}}=\dot{\Gamma}/\Gamma$, where $\Gamma\equiv b_xb_yb_z$ is the volume scale factor and $\Gamma$ and $\sigma_{ii}$ are functions only of the time. The scale factors obey
\begin{equation}
\ddot{b}_i=\frac{\overline{\omega_i^2}}{\Gamma^{2/3}b_i}[1\,+\,C(t)]
-\frac{\hbar\langle\alpha_S\rangle\,\sigma_{ii}}{m\langle x_i^2\rangle_0\, b_i}-\omega^2_{imag}b_i.
\label{eq:6.1e}
\end{equation}
In Eq.~\ref{eq:6.1e}, we define $\overline{\omega_i^2}\equiv\widetilde{E}/(3m\langle x_i^2\rangle_0)$ for an arbitrary trapping potential, which need not be harmonic~\cite{SupportOnline} and $\omega^2_{ymag}=\omega^2_{zmag}=\omega^2_{mag}$ and $\omega^2_{xmag}=-2\,\omega^2_{mag}$ (repulsive), with $\omega_{mag}$ defined above.

The coefficient $C(t)=C_Q(t)+C_{\Delta p}(t)$ in Eq.~\ref{eq:6.1e}  includes two effects exactly (within the scaling approximation): $C_Q$ is the fractional increase in the volume integrated pressure arising from viscous heating, which is determined by $\dot{Q}$~\cite{SupportOnline}. $C_{\Delta p}(t)$ describes the corresponding fractional change for a given conformal symmetry breaking pressure change $\Delta p(t)$. For the transverse aspect ratio, $\sigma_x/\sigma_y$, we find that $C_Q$ is important, but that $C_{\Delta p}$ has a negligible effect~\cite{SupportOnline}.

The shear viscosity coefficient $\langle\alpha_S\rangle$ is measured by using Eq.~\ref{eq:6.1e} to fit the data for the transverse aspect ratio $\sigma_x/\sigma_y=\omega_{y}b_x/(\omega_{x}b_y)$ as a function of time after release, while self consistently determining $\widetilde{E}/E_F$ from the measured cloud sizes  $\sigma_x, \sigma_y, \sigma_z,$ and $N$. At resonance, where the scattering length $a$ diverges and $1/(k_{FI}a)=0$, $\alpha_S$ can be a function only of the local reduced temperature $\theta\propto T/n^{2/3}$. As the viscosity makes a small perturbation to the flow, we approximate the temperature within the viscosity coefficient in zeroth order, i.e., we assume that the temperature evolves adiabatically during the expansion after  the optical trap is abruptly extinguished. Then $T\propto n^{2/3}$, so that the local $\theta$ remains fixed at its initial value. In this case, $\langle\alpha_S\rangle\equiv\langle\alpha_S\rangle_0$ is temporally constant as the cloud expands, i.e., it is equal to the trap-averaged initial value with $n\rightarrow n_0$.

\begin{figure}[htb]
\includegraphics[width=3.5in]{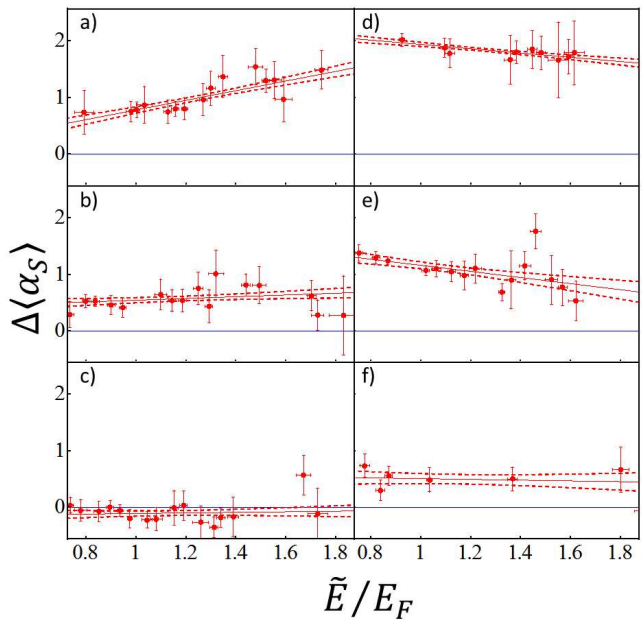}
\caption{Difference in shear viscosity on and off resonance, $\Delta\langle\alpha_S\rangle$, versus energy $\widetilde{E}/E_F$ and interaction strength $1/(k_{FI} a)$.  On the BEC side of resonance (left column),  $1/(k_{FI} a)=$ a) $0.83(6)$, b) $0.55(5)$, c) $0.25(3)$. On the BCS side of resonance (right column) $1/(k_{FI}a)=$ d) $-0.61(1)$, e) $-0.34(1)$, f) $-0.16(3)$. Red line denotes linear fit to $\Delta\langle\alpha_S\rangle(\widetilde{E})$. Dotted red lines show 1-$\sigma$ confidence interval.  Blue zero lines show the unitary fit to shear viscosity $\langle\alpha_S\rangle_0$ by construction. \label{fig:difference}}
\end{figure}

We determine $\langle\alpha_S\rangle$ both on resonance, where $\langle\alpha_S\rangle = \langle\alpha_S\rangle_0$ is temporally constant, and at finite $1/(k_{FI}a)$, initially ignoring the time dependence arising from the finite scattering length, which we include later in Eq.~\ref{eq:shear}. Fig.~\ref{fig:difference} shows the  difference $\Delta\langle\alpha_S\rangle=\langle\alpha_S\rangle-\langle\alpha_S\rangle_0$  between the $\langle\alpha_S\rangle$ determined at finite $1/(k_{FI}a)$ and the resonant value $\langle\alpha_S\rangle_0$. We determine $\langle\alpha_S\rangle_0$ from a polynomial fit to the resonant shear viscosity as a function of $\widetilde{E}/E_F$~\cite{SupportOnline}.

We find that the minimum in shear viscosity occurs on BEC side of resonance, Fig.~\ref{fig:difference}c. Further, we note that the $\Delta\langle\alpha_S\rangle$ depends strongly on the magnitude and sign of the interaction strength $1/(k_{FI}a)$.  Generally, on the BEC side of resonance $1/(k_{FI}a)>0$, we find that $\Delta\langle\alpha_S\rangle$ increases with with increasing energy, which may arise from a corresponding decrease in the dimer fraction, as discussed further below.  On the BCS side of resonance $1/(k_{FI}a)<0$, $\Delta\langle\alpha_S\rangle$ decreases with  increasing energy, which may arise from reduced Pauli blocking, i.e., the collision rate increases with temperature in the degenerate regime.  Clearly, a simple quadratic dependence on the interaction strength is insufficient to encompass all the observed behavior of $\Delta\langle\alpha_S\rangle$.

\begin{figure}[htb]
\includegraphics[width=3.5in]{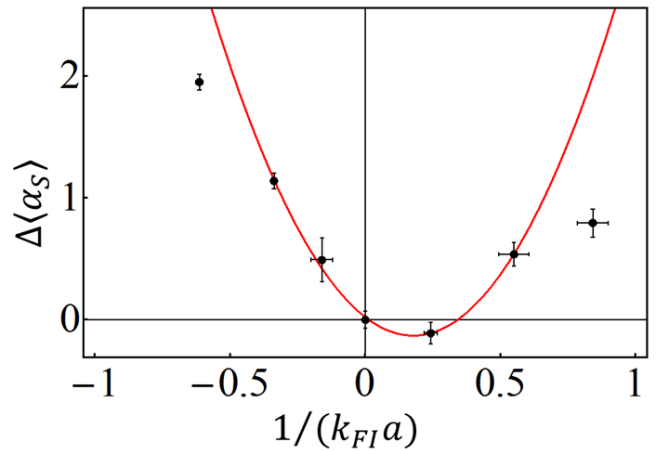}
\caption{Difference in shear viscosity on and off resonance, $\Delta\langle\alpha_S\rangle$, versus interaction strength $1/(k_{FI} a)$ at an energy of $\widetilde{E}/E_F = 1$. Black circles represent $\Delta\langle\alpha_S\rangle$ obtained from the linear fits in Fig.~\ref{fig:difference}.  Vertical errorbars are the 1-$\sigma$ confidence interval of the fits.  Red line is best fit of $\tilde{c}_0+\tilde{c}_1/(k_{FI}a)+\tilde{c}_2/(k_{FI}a)^2$ to the data with $\tilde{c}_0=0.0$, $\tilde{c}_1=-1.7$, and $\tilde{c}_2=4.8$. From the fit, the minimum occurs at $1/(k_F a) = 0.18$.
\label{fig:deltaalphavskFa}}
\end{figure}

In order to investigate further, we fit a linear energy dependence to $\Delta\langle\alpha_S\rangle$ for each interaction strength $1/(k_{FI}a)$ as shown in Fig.~\ref{fig:difference}.  In Fig.~\ref{fig:deltaalphavskFa}, $\Delta\langle\alpha_S\rangle$ is plotted as a function of $1/(k_{FI}a)$ for a fixed energy $\widetilde{E}/E_F = 1$. We see a nominally parabolic dependence on $1/(k_{FI}a)$, with the minimum clearly shifted toward the BEC side of resonance. Setting $\Delta\langle\alpha_S\rangle=\tilde{c}_0+\tilde{c}_1/(k_{FI}a)+\tilde{c}_2/(k_{FI}a)^2$, we fit the data shown in Fig.~\ref{fig:difference} excluding the two extreme $1/(k_{FI}a)$ points where a simple perturbation expansion in $1/(k_{FI} a)$ is likely to break down.  We find $\tilde{c}_0=0.0$, $\tilde{c}_1=-1.7$, and $\tilde{c}_2=4.8$.  Recall that we have ignored the expansion time dependence arising from the finite scattering length.  Therefore, we can only draw qualitative conclusions from our fit to $\Delta\langle\alpha_S\rangle$ versus $1/(k_{FI} a)$.

We now obtain quantitative results for the dependence of the shear viscosity on $1/(k_{FI}a)$ by including the explicit time dependence of the shear viscosity coefficients and refitting the data.
Using dimensional analysis, the leading-order scattering-length-dependent terms in local shear viscosity take the forms $\hbar n\,f_1(\theta)/(k_F a)$ $\hbar n\,f_2(\theta)/(k_F a)^2$, where $f_{1,2}(\theta)$ are dimensionless functions of the reduced temperature. Then, in the scaling approximation described above, the density $n\propto k_F^3$ decreases by the volume scale factor $\Gamma$ as the cloud expands, so that $1/k_F\propto\Gamma^{1/3}(t)$. For the viscosity coefficients, we again approximate the temperature to zeroth order as evolving adiabatically, so that  $f_{1,2}(\theta)$ are temporally constant.  Averaging over the cloud volume, as in Eq.~\ref{eq:trapavcoeff}, we then obtain the general form for the time-dependent cloud-averaged viscosity coefficient,
\begin{equation}
\langle\alpha_S\rangle = \langle\alpha_S\rangle_0+c_1\frac{\Gamma^{1/3}(t)}{k_{FI}a}\,+c_2\frac{\Gamma^{2/3}(t)}{(k_{FI}a)^2}.
\label{eq:shear}
\end{equation}
In the spirit of a perturbation expansion in $1/(k_{FI}a)$ about resonance at fixed $\widetilde{E}$, the first term is taken to be the shear viscosity coefficient at resonance, which is time-independent and determined versus $\widetilde{E}$, as described above.

\begin{figure}[htb]
\includegraphics[width=3.5in]{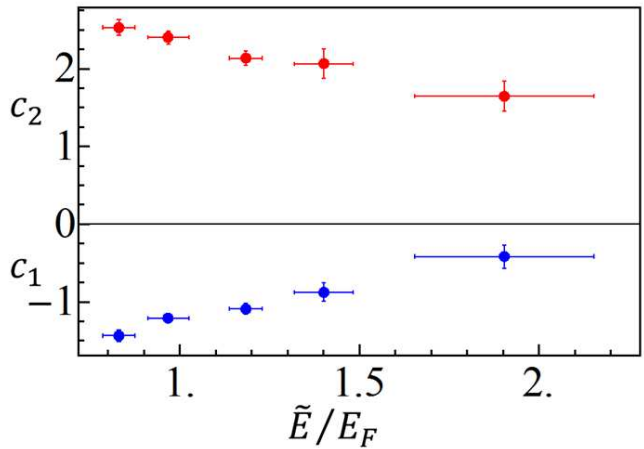}
\caption{Off resonant shear viscosity coefficients $c_1$ and $c_2$ from Eq.~\ref{eq:shear}, $\langle\alpha_S\rangle_0+c_1\frac{\Gamma^{1/3}(t)}{k_{FI}a}\,+c_2\frac{\Gamma^{2/3}(t)}{(k_{FI}a)^2}$,
 obtained by integrating Eq.~\ref{eq:6.1e} and globally minimizing $\chi^2$ for aspect ratio data within a range of energies and interactions strengths.  Red (upper) circles are $c_2$ versus energy.  Blue (lower) circles are $c_1$ versus energy. \label{fig:ChisSuareResults}
 }
\end{figure}

We globally fit the data over discrete energy ranges and limit the  range of interaction strength to $-0.5 < 1/(k_{Fi} a) < 0.7$. This is accomplished by integrating Eq.~\ref{eq:6.1e}, using  Eq.~\ref{eq:shear}. As $\langle\alpha_S\rangle_0$ is known as a function of $\widetilde{E}$,  $c_1$ and $c_2$ are used as fit parameters, determined by a $\chi ^2$ fit to the aspect ratio data. The results are shown in Fig.~\ref{fig:ChisSuareResults}.

For an energy range of $0.9 < \widetilde{E}/E_F < 1.1$, $71$ points are included in the two parameter fit with an average energy of $\widetilde{E}/E_F = 0.97 (6)$.  We obtain a normalized $\chi^2 = 1.1$, with $c_1 = -1.22(5)$ and $c_2 = 2.43(9)$. Eq.~\ref{eq:shear} then yields a minimum in the initial (in-trap) shear viscosity at $1/(k_{FI} a) = -c_1/(2c_2) = 0.25$.   For an energy of $\widetilde{E}/E_F = 0.97$, the polynomial fit for the resonant gas gives a shear viscosity of $\langle\alpha_S\rangle_0 = 1.10$, so that  at $t=0$, $\langle\alpha_S\rangle_{min}=\langle\alpha_S\rangle_0-c_1^2/(4c_2)=0.95$. Note that the smaller values of the in-trap $c_{1,2}$ compared to $\tilde{c}_{1,2}$ are consistent with the time-dependent factors in Eq.~\ref{eq:shear}, since $\Gamma(t)$ increases from unity as the cloud expands.  Complete results can be found in Ref.~\cite{SupportOnline}.

The shift of the minimum shear viscosity toward the BEC side of resonance may be explained by an enhancement in the bosonic degrees of freedom~\cite{LevinViscosity}, such as preformed atom pairs or dimer molecules. These bosonic degrees of freedom would suppress Pauli blocking and increase the effective scattering rate~\cite{BruunViscousNormalDamping}. In addition, the collisional cross section for dimer-atom scattering is larger than that for atom-atom scattering~\cite{PetrovAtomDimerScattLength}. Therefore, since the shear viscosity scales inversely with the scattering rate, one would expect the observed decrease on the BEC side.  We note also that $\langle\alpha_S\rangle$ is well below the parabolic fit at the two extreme points $1/(k_{FI}a)=0.83$ and $1/(k_{FI}a)=-0.61$. This may be a consequence of a divergence of the expansion in  powers $1/(k_{FI}a)$, but may also be the result of a larger dimer fraction on the BEC side of resonance.  We observe also that the location of the minimum in the shear viscosity, $-c_1/(2c_2)$, moves toward resonance with increasing energy~\cite{SupportOnline}, indicating that the $\langle\alpha_S\rangle$ may scale quadratically with $1/(k_{FI}a)$ at higher temperatures, as predicted~\cite{BruunViscosity2}.

This research is supported by the Physics Division of the National Science Foundation (quantum transport in strongly interacting Fermi gases PHY-1067873), the Division of Materials Science and Engineering,  the Office of Basic Energy Sciences, Office of Science, U.S. Department of Energy (thermodynamics in strongly correlated Fermi gases DE-SC0008646), the Physics Divisions of the Army Research Office (strongly interacting Fermi gases in reduced dimensions W911NF-11-1-0420) and the Air Force Office of Scientific Research (non-equilibrium Fermi gases FA9550-13-1-0041).  The authors are pleased to acknowledge K. Dusling and T. Sch\"{a}fer, North Carolina State University, for stimulating conversations.

%\bibliography{Scaling-4-2014BibTexData}

\newpage
\widetext
\section{Supplemental Material}
\label{sec:supplement}

In this supplemental material, we derive new evolution equations  for the radii of a hydrodynamically expanding cloud as a function of interaction strength near a collisional (Feshbach) resonance.  The evolution equations, which are based on a scaling approximation, include the change in the pressure $\Delta p$, relative to the resonant regime, which breaks conformal symmetry, as well as the forces and heating arising from viscosity. We justify a single component hydrodynamic description by estimating the Knudsen number and investigating the transition from hydrodynamic to ballistic flow as the cloud expands. We show that by measuring the transverse aspect ratio $\sigma_x/\sigma_y$, as a function of time after release of the cloud, we are able to precisely determine the shear viscosity. We also show that $\sigma_x/\sigma_y$ is insensitive to both $\Delta p$ and the bulk viscosity, enabling a single parameter fit to the transverse aspect ratio data to determine the shear viscosity coefficient. Finally, we provide a detailed description of the data analysis and the complete results.

\subsection{Hydrodynamic Theory}
\label{sec:hydro}
To determine the viscosity on and away from the Feshbach resonance, we employ a hydrodynamic description for a single-component fluid. The single fluid description is justified in Ref.~\cite{StringariBulk} for the expansion of the resonantly interacting gas in the normal fluid regime and below the superfluid transition where the normal and superfluid components expand together.  As noted in the main text, for our experiments below resonance, a finite dimer pair fraction can exist. At low temperatures, the dimers can condense into a BEC and a two-fluid description would be required.  We avoid this complication by working in the normal fluid regime, where we do not observe a condensate fraction.  Further justification for the single component hydrodynamic description based on the Knudsen number can be found in Sec.\ref{sec:Assumptions}.

\subsubsection{Hydrodynamic Equations}
For a single component fluid, the velocity field $\mathbf{v}(\mathbf{r},t)$ is determined by the scalar pressure and the viscous stress tensor,
\begin{equation}
n\,m\left(\partial_t +\mathbf{v}\cdot\nabla\right)v_i=-\partial_i p + \sum_j \partial_j (\eta\,\sigma_{ij}+\zeta_B\,\sigma^{'}\delta_{ij})-n\,\partial_i U_{total}.
\label{eq:force}
\end{equation}
Here $p$ is the scalar pressure and $m$ is the atom mass.  $U_{total}$ is the total trapping potential energy arising from the optical trap $U_{opt}$ and the bias magnetic field curvature $U_{mag}$, as described in the main text.  The second term on the right describes the friction forces
arising from both shear $\eta$ and bulk $\zeta_B$ viscosities, where $\sigma_{ij}=\partial
v_i/\partial x_j+\partial v_j/\partial
x_i-2\delta_{ij}\nabla\cdot\mathbf{v}/3$  and $\sigma^{'}\equiv\nabla\cdot\mathbf{v}$. Current conservation for the density $n({\mathbf{r}},t)$ requires
\begin{equation}
\frac{\partial n}{\partial t}+\nabla\cdot(n{\mathbf{v}})=0.
\label{eq:ncons}
\end{equation}
Finally,  consistent with Eq.~\ref{eq:force} and Eq.~\ref{eq:ncons}, conservation of the total energy  is described by
\begin{equation}
\frac{d}{d t}\int d^3{\mathbf{r}}\left(n\frac{1}{2}m{\mathbf{v}}^2+{\cal E}+n\,U_{total}\right)=0.
\label{eq:energycons}
\end{equation}
The first term in Eq.~\ref{eq:energycons} is the kinetic energy of the velocity field and ${\cal E}$ is the internal energy density of the cloud.

For each direction $i=x,y,z$, the mean square size $\langle x_i^2\rangle\equiv\frac{1}{N}\int d^3{\mathbf{r}}\,n({\mathbf{r}},t)\,x_i^2$ obeys
\begin{eqnarray}
\frac{d\langle x_i^2\rangle}{d t}&=&\frac{1}{N}\int d^3{\mathbf{r}}\,\frac{\partial n}{\partial t}x_i^2=\frac{1}{N}\int d^3{\mathbf{r}}\,[-\nabla\cdot(n{\mathbf{v}})]x_i^2=\frac{1}{N}\int d^3{\mathbf{r}}\,n\,{\mathbf{v}}\cdot\nabla x_i^2\nonumber\\
&=&2\langle x_i\,v_i\rangle,
\label{eq:xsq}
\end{eqnarray}
where $N$ is the total number of atoms.  We have used integration by parts  and $n=0$ for $x_i\rightarrow\pm\infty$ to obtain the second line. Here, and throughout the discussion, $\langle ...\rangle\equiv\int d^3{\mathbf{r}}\,(...) \,n({\mathbf{r}},t)/N$ denotes the cloud average with respect to the normalized density.  Similarly,
\begin{eqnarray}
\frac{d\langle x_iv_i\rangle}{d t}&=&\frac{1}{N}\int d^3{\mathbf{r}}\,n\,x_i\frac{\partial v_i}{\partial t}+\frac{1}{N}\int d^3{\mathbf{r}}\,\frac{\partial n}{\partial t}\,x_i v_i=\frac{1}{N}\int d^3{\mathbf{r}}\,n\,x_i\frac{\partial v_i}{\partial t}+\frac{1}{N}\int d^3{\mathbf{r}}\,n\,{\mathbf{v}}\cdot\nabla (x_iv_i)\nonumber\\
&=&\langle x_i(\partial_t+{\mathbf{v}}\cdot\nabla)v_i\rangle+\langle v_i^2\rangle.
\label{eq:xv}
\end{eqnarray}
Combining Eq.~\ref{eq:xsq} and Eq.~\ref{eq:xv}, we obtain,
\begin{equation}
\frac{d^2}{d t^2}\frac{\langle x_i^2\rangle}{2}=\langle x_i(\partial_t+{\mathbf{v}}\cdot\nabla)v_i\rangle+\langle v_i^2\rangle.
\label{eq:xsqddot1}
\end{equation}
To proceed, we use Eq.~\ref{eq:force}, which yields
\begin{equation*}
\int d^3{\mathbf{r}}\,n\,x_i(\partial_t+{\mathbf{v}}\cdot\nabla)v_i= \frac{1}{m}\int d^3{\mathbf{r}}\,x_i(-\partial_i p-n\,\partial_i U_{total})+\frac{1}{m}\sum_j\int d^3{\mathbf{r}}\,x_i\partial_j (\eta\,\sigma_{ij}+\zeta_B\,\sigma^{'}\delta_{ij})
\end{equation*}
Integrating by parts on the right hand side, assuming that the surface terms vanish, we obtain
\begin{equation}
\langle x_i(\partial_t+{\mathbf{v}}\cdot\nabla)v_i\rangle=\frac{1}{Nm}\int d^3{\mathbf{r}}\,p-\frac{1}{m}\langle x_i\partial_iU_{total}\rangle-\frac{1}{Nm}\int d^3{\mathbf{r}}\,(\eta\,\sigma_{ii}+\zeta_B\,\sigma')
\end{equation}
with $\sigma'\equiv\nabla\cdot{\mathbf{v}}$.
Using $\hbar\,n$ as the natural scale of viscosity, we define the shear and bulk viscosity coefficients $\alpha_S$ and $\alpha_B$ by $\eta\equiv\alpha_S\,\hbar\,n$ and $\zeta_B\equiv\alpha_B\,\hbar\,n$, respectively. Then,
\begin{equation}
\langle x_i(\partial_t+{\mathbf{v}}\cdot\nabla)v_i\rangle=\frac{1}{Nm}\int d^3{\mathbf{r}}\,p-\frac{1}{m}\langle x_i\partial_iU_{total}\rangle-\frac{\hbar}{m}\langle\alpha_S\,\sigma_{ii}+\alpha_B\,\sigma'\rangle,
\label{eq:2.4a}
\end{equation}
where
\begin{equation}
\langle\alpha_S\,\sigma_{ii}+\alpha_B\,\sigma'\rangle\equiv\frac{1}{N}\int d^3{\mathbf{r}}\,n\,(\alpha_S\,\sigma_{ii}+\alpha_B\,\sigma').
\end{equation}
Using Eq.~\ref{eq:2.4a} in Eq.~\ref{eq:xsqddot1}, we then obtain for one direction $x_i$,
\begin{equation}
\frac{d^2}{dt^2}\frac{\langle x_i^2\rangle}{2}=\frac{1}{Nm}\int d^3{\mathbf{r}}\,p+\langle v_i^2\rangle-\frac{1}{m}\langle x_i\partial_i U_{total}\rangle-\frac{\hbar}{m}\langle\alpha_S\,\sigma_{ii}+\alpha_B\,\sigma'\rangle.
\label{eq:xsqddot2}
\end{equation}
Eq.~\ref{eq:xsqddot2} determines the evolution of the mean square cloud radii along each axis, $\langle x_i^2\rangle$, which depends on the conservative forces arising from the scalar pressure and the trap potential, as well as the viscous forces arising from the shear and bulk viscosities.

\subsection{Scaling Solution}
\label{sec:scaling}
We determine the viscosity by measuring the cloud radii and the transverse aspect ratio as a function of time after the cloud is released from the trap. To analyze the aspect ratio data, we employ a scaling solution of Eq.~\ref{eq:xsqddot2}, where the density is given by
\begin{equation}
n({\mathbf{r}},t)=\frac{n_0(x/b_x,y/b_y,z/b_z)}{\Gamma},
\label{eq:density}
\end{equation}
where $b_i(t)$, $i=x,y,z$ is a time dependent scale factor, with $b_i(0)=1$ and $\dot{b}_i(0)=0$. $n_0$ is the  density profile of the trapped cloud in equilibrium. Here, $\Gamma(t)\equiv b_xb_yb_z$ is the volume scale factor, which is independent of the spatial coordinates in the scaling approximation. With Eq.~\ref{eq:density} and a velocity field that is linear in the spatial coordinates, $v_i=x_i\,\dot{b}_i/b_i$, Eq.~\ref{eq:ncons} is automatically satisfied.

We note that $\langle x_i^2\rangle=\langle x_i^2\rangle_0\,b_i^2(t)$, and $\langle v_i^2\rangle=\langle x_i^2\rangle\,\dot{b_i}^2/b_i^2=\langle x_i^2\rangle_0\,\dot{b_i}^2(t)$, where $\langle x_i^2\rangle_0$ is the mean-square cloud radius of the trapped cloud in the $i^{th}$ direction, just before release. Then, with these scaling assumptions, Eq.~\ref{eq:xsqddot2} yields
\begin{equation}
\langle x_i^2\rangle_0\,b_i\,\ddot{b}_i=\frac{1}{Nm}\int d^3{\mathbf{r}}\,p-\frac{1}{m}\langle x_i \partial_i U_{total}\rangle-\frac{\hbar}{m}\langle\alpha_S\,\sigma_{ii}+\alpha_B\nabla\cdot{\mathbf{v}}\rangle.
\label{eq:2.4}
\end{equation}

We see that Eq.~\ref{eq:2.4} contains the pressure only in a volume integral. To determine the evolution equation for the pressure integral, we use energy conservation. We begin by defining
\begin{equation}
\Delta p\equiv p-\frac{2}{3}\,{\cal E}.
\label{eq:deltap}
\end{equation}
As noted in our previous study of scale invariance~\cite{ElliottScaleInv}, $\Delta p$ is the conformal symmetry breaking pressure change, which vanishes at resonance, where $p=\frac{2}{3}\,{\cal E}$.  Analogous to the methods used to derive Eq.~\ref{eq:xsqddot1}, we move the time derivatives of the velocity field and density inside the integral in Eq.~\ref{eq:energycons} and use Eq.~\ref{eq:force} and Eq.~\ref{eq:ncons} to obtain an evolution equation for the volume integral of the energy density,
\begin{equation}
\frac{d}{d t}\int d^3{\mathbf{r}}\,{\cal E}+\int d^3{\mathbf{r}}(\nabla\cdot{\mathbf{v}})\,p\, +
\int d^3{\mathbf{r}}\,n\,\frac{\partial U_{total}}{\partial t}=\dot{Q},
\label{eq:3.1e}
\end{equation}
where $\dot{Q}\equiv\int d^3{\mathbf{r}}\,\dot{q}$ is the total heating rate arising from the friction forces and $\dot{q}$ is the heating rate per unit volume,
\begin{equation}
\dot{q}=\frac{1}{2}\eta\sum_{ij}\sigma_{ij}^2+\zeta_B(\nabla\cdot{\mathbf{v}})^2.
\label{eq:2.1}
\end{equation}
Just after release of the cloud, the trap potential is constant in time and  $\partial_t U_{total}$ vanishes. Then for each volume element $d^3{\mathbf{r}}$, Eq.~\ref{eq:3.1e} is just $dE_{int}=dQ-p\,dV$, where $dE_{int}$ is the internal energy in the volume element, $dQ$ is the heat added to the volume element in a time $dt$, and the work done  by the volume element arises from expansion, $dV=d^3{\mathbf{r}}\,(\nabla\cdot{\mathbf{v}})\,dt$.

Using $\partial_t U_{total}=0$ and Eq.~\ref{eq:deltap} to eliminate ${\cal E}$, Eq.~\ref{eq:3.1e} takes the form
\begin{equation}
\frac{d}{d t}\int d^3{\mathbf{r}}\,p+\frac{2}{3}\int d^3{\mathbf{r}}\,(\nabla\cdot{\mathbf{v}})\,p=\frac{2}{3}\,\dot{Q}+\frac{d}{dt}\int d^3{\mathbf{r}}\,\Delta p.
\label{eq:3.2e}
\end{equation}
As we intend to explore small deviations from the scale invariant regime, the last term on the right of Eq.~\ref{eq:3.2e} can be evaluated using suitable approximations, as discussed below. It vanishes if the volume integral of $\Delta p$ is time independent.

With the scaling assumptions,  $\nabla\cdot{\mathbf{v}}=\dot\Gamma/\Gamma$ is independent of the spatial coordinates, and Eq.~\ref{eq:3.2e} reduces to
\begin{equation}
\frac{d}{d t}\int d^3{\mathbf{r}}\,p+\frac{2}{3}\frac{\dot{\Gamma}}{\Gamma}\,\int d^3{\mathbf{r}}\,p=\frac{2}{3}\dot{Q}+\frac{d}{dt}\int d^3{\mathbf{r}}\,\Delta p.
\label{eq:2.9e}
\end{equation}

Using the integrating factor $\Gamma^{2/3}$, integration of Eq.~\ref{eq:2.9e} from $t=0$ to $t$ yields
\begin{equation}
3\Gamma^{2/3}\int d^3{\mathbf{r}}\,p=3\int d^3{\mathbf{r}}\,p_0+2\int_0^t dt\,\Gamma^{2/3}\dot{Q}+3\left[\Gamma^{2/3}\int d^3{\mathbf{r}}\,\Delta p-\int d^3{\mathbf{r}}\,\Delta p_0\right]-
2\int_1^\Gamma\frac{d\Gamma}{\Gamma^{1/3}}\int d^3{\mathbf{r}}\,\Delta p,
\label{eq:3.4e}
\end{equation}
where we have used $\Gamma(0)=1$. Here $p_0$ and $\Delta p_0$ denote the initial pressure and the conformal symmetry breaking pressure just after release.

Eq.~\ref{eq:3.4e} is a general consequence of energy conservation. Although it can be used to determine the evolution of $\int d^3{\mathbf{r}}\,p$ in general, it is particularly well-suited to a perturbative treatment of $\Delta p$ in the near scale-invariant regime. In that case,  we can approximate the time-dependence of the temperature in $\Delta p$ as adiabatic, i.e., $T=T_0\,\Gamma^{-2/3}$, where $T_0$ is the initial temperature of the trapped cloud. Then the volume integral of $\Delta p$ becomes a known function of time, as discussed below in more detail in \S~\ref{sec:highT}.

We find the initial condition $\int d^3{\mathbf{r}}\,p_0$ from
\begin{equation}
\frac{3}{N}\int d^3{\mathbf{r}}\,p_{\,0}=\langle {\mathbf{r}}\cdot\nabla U_{total}\rangle_0\equiv\widetilde{E},
\label{eq:1.2}
\end{equation}
with the  energy scale $\widetilde{E}$ defined and measured as described in the main text.

Using  Eq.~\ref{eq:3.4e}, we write the time-dependent volume integral of the pressure using  Eq.~\ref{eq:3.4e} in the form,
\begin{equation}
\frac{1}{N}\int d^3{\mathbf{r}}\,p=\frac{\langle{\mathbf{r}}\cdot\nabla U_{total}\rangle_0}{3\,\Gamma^{2/3}}\left[1+ C_Q(t)+ C_{\Delta p}(t)\right].
\label{eq:pressint}
\end{equation}
Here, the fractional change in the pressure integral due to viscous heating is given by $C_Q(t)$, which is determined from
\begin{equation}
\dot{C}_Q(t)\equiv\frac{\Gamma^{2/3}(t)\frac{2\dot{Q}}{N}}{\langle{\mathbf{r}}\cdot\nabla U_{total}\rangle_0},
\label{eq:6.2e}
\end{equation}
with the initial condition $C_Q(0)=0$. Using Eq.~\ref{eq:2.1} with the velocity field $v_i=x_i\,\dot{b}_i/b_i$, where $\partial_j v_i=\delta_{ij}\dot{b}_i/b_i$ is spatially constant, it is straightforward to obtain
\begin{equation}
\frac{2\dot{Q}}{N}=\hbar\,\langle \alpha_S \rangle\,\sum_i\sigma_{ii}^2+2\hbar\,\langle \alpha_B \rangle\,
\frac{\dot{\Gamma}^2}{\Gamma^2}.
\label{eq:heat2}
\end{equation}
 The trap averaged-viscosity coefficients, which appear in Eq.~\ref{eq:heat2}, are defined by
 \begin{eqnarray}
 \langle \alpha_S \rangle&\equiv&\int d^3{\mathbf{r}}\,\eta/(N\hbar)\nonumber\\
 \langle \alpha_B \rangle&\equiv&\int d^3{\mathbf{r}}\,\zeta_B/(N\hbar).
 \label{eq:viscoeff}
 \end{eqnarray}
 In general, the trap-averaged viscosity coefficients are dependent on the scattering length $a$ and are time-dependent, as described in the main text.

 In Eq.~\ref{eq:heat2},
 \begin{equation}
 \frac{\dot{\Gamma}}{\Gamma}=\frac{\dot{b}_x}{b_x}+\frac{\dot{b}_y}{b_y}+\frac{\dot{b}_z}{b_z}
 \end{equation}
 and
 \begin{equation}
 \sigma_{ii}=2\frac{\dot{b}_i}{b_i}-\frac{2}{3}\frac{\dot{\Gamma}}{\Gamma}.
 \end{equation}
Then,
\begin{equation}
\sum_i\sigma_{ii}^2=4\sum_i\frac{\dot{b}_i^2}{b_i^2}-\frac{4}{3}\frac{\dot{\Gamma}^2}{\Gamma^2}.
\end{equation}

The time-dependent $\Delta p$ terms in Eq.~\ref{eq:3.4e}  give the net fractional change in the pressure integral arising from  the conformal symmetry breaking pressure $\Delta p$,
\begin{equation}
C_{\Delta p}(t)\equiv C_F(t)-C_F(0)-C_p(t),
\label{eq:Cdeltap}
\end{equation}
where
\begin{equation}
C_F(t)\equiv \frac{\Gamma^{2/3}(t)\frac{3}{N}\int d^3{\mathbf{r}}\,\Delta p}{\langle{\mathbf{r}}\cdot\nabla U_{total}\rangle_0}.
\label{eq:6.4e}
\end{equation}
and
\begin{equation}
C_p(t)\equiv\frac{2\int_1^{\Gamma(t)}\frac{d\Gamma}{\Gamma^{1/3}}\frac{1}{N}\int d^3{\mathbf{r}}\,\Delta p}{\langle{\mathbf{r}}\cdot\nabla U_{total}\rangle_0}.
\label{eq:6.5e}
\end{equation}
From Eq.~\ref{eq:6.4e} and Eq.~\ref{eq:6.5e}, we see that $C_{\Delta p}$ of Eq.~\ref{eq:Cdeltap} vanishes at $t=0$, and also when $\Delta p$ is time independent, as it should.

With Eq.~\ref{eq:pressint} for the volume integral of the pressure,  Eq.~\ref{eq:2.4} yields our central result for the scale factor evolution,
\begin{equation}
\ddot{b}_i=\frac{\overline{\omega_i^2}}{\Gamma^{2/3}b_i}\left[1+ C_Q(t)+ C_{\Delta p}(t)\right]-\frac{\hbar\left(\langle \alpha_S \rangle\,\sigma_{ii}
+\langle \alpha_B \rangle\,\frac{\dot{\Gamma}}{\Gamma}\right)}{m\langle x_i^2\rangle_0 b_i}-\frac{\langle x_i \partial_i U_{mag}\rangle}{m\langle x_i^2\rangle_0 b_i}.
\label{eq:6.1e}
\end{equation}

In the last term of Eq.~\ref{eq:6.1e}, note that $U_{total}$ is replaced by the magnetic potential, $U_{mag}$ defined in the main text, as we are interested in expansion of the cloud after the optical part of the potential is extinguished.
Further, we have defined the mean square ballistic frequency for an arbitrary trapping potential, which need not be harmonic,
\begin{equation}
\overline{\omega_i^2}\equiv\frac{\langle x_i\partial_i U_{total}\rangle_0}{m\langle x_i^2\rangle_0}
=\frac{\langle{\mathbf{r}}\cdot\nabla U_{total}\rangle_0}{3m\langle x_i^2\rangle_0}.
\label{eq:4.6e}
\end{equation}
Here, $U_{total}$ is the total trap potential {\it prior} to release of the cloud. The second form follows from force balance  in equilibrium, $\partial_i p+n\partial_i U_{total}=0$. Multiplying by $x_i$ and integrating by parts requires that $\langle x_i\partial_i U_{total}\rangle_0$ be the same for all directions. We determine $\overline{\omega_i^2}$ from the measured cloud profile and trap parameters, which are given in the main text.

Eq.~\ref{eq:6.1e}  determines the expansion factors $b_i$ with the initial conditions $b_i(0)=1$ and $\dot{b}_i(0)=0$, using the known trap parameters and a suitable approximation for $\Delta p$ in the off-resonance case. The trap-averaged (generally time-dependent) shear and bulk viscosity coefficients, $\langle \alpha_S \rangle$ and $\langle \alpha_B \rangle$ are used as fit parameters, as described in the main text. In the experiments, we determine $\langle \alpha_S \rangle$  by fitting the predicted aspect ratios to the aspect ratio data,  neglecting the much smaller $\langle \alpha_B \rangle$. The bulk viscosity coefficient $\langle \alpha_B \rangle$  is measured by observing the mean square cloud radius $\langle{\mathbf{r}}^2\rangle$, which is a scalar,  as a function of time after release, as described in Ref.~\cite{ElliottScaleInv}.

\subsection{Basic Assumptions}
\label{sec:Assumptions}
Here we elucidate the basic assumptions underlying the data analysis.  First, we address the question of the validity of the hydrodynamic model both on and off resonance.  Next we show that for measuring the shear viscosity, we can ignore the effects of the bulk viscosity $\langle \alpha_B \rangle$ and the conformal symmetry breaking pressure $\Delta p$. Qualitatively, these scalar parameters uniformly slow or accelerate the expansion, affecting each direction in the same way. Hence, the change in the aspect ratio $\sigma_x/\sigma_y$ is suppressed. In contrast, the transverse aspect ratio $\sigma_x/\sigma_y$ is very sensitive to the shear viscosity, which  directs momentum from the more rapidly expanding direction $x$ into the less rapidly expanding direction $y$.

\subsubsection{Knudsen Number and Validity of a Hydrodynamic Model}

To investigate a possible breakdown of hydrodynamics, we consider the high temperature regime, where the system is most likely to deviate from hydrodynamic flow, since the initial density $n$ and collision cross section $\sigma$ are smallest. Assuming that the cloud comprises a normal fluid mixture of atoms and dimers, the gas will be hydrodynamic and will expand as a single fluid if the Knudsen number $K_n$ is small for all species. Using a classical scattering description, we take for $K_n$  the  ratio of the collisional mean free path $\lambda_{mfp}$ to the smallest diameter of the cloud $2R_x$, analogous to our previous treatment~\cite{NJPReview}. When the Knudsen number for all species is small, one expects that the  dimer component of the cloud will move together with the atom component. In this case, a single-component hydrodynamic description is valid. As first shown in Ref.~\cite{PetrovDimerDimerScattLength}, for a two-component Fermi gas near a Feshbach resonance, the dimer-dimer scattering length is $0.6\,a$, where $a$ is the scattering length for a collision between a spin-up and a spin-down atom.  For dimer-atom collisions, the scattering length is $1.2\,a$~\cite{PetrovAtomDimerScattLength}. We expect a small dimer fraction at high temperature, so that dimer-atom collisions are predominant in determining the hydrodynamic behavior of the dimers. Since the dimer-atom scattering cross section is larger than that for atom-atom scattering and the dimers scatter from both atomic species, i.e., from the  total atomic density, the Knudsen number for the dimers will be smaller than that for the atoms. Hence, a conservative estimate of the relevant Knudsen number can be based on the  mean free path for the atom component.

We assume that the dimer fraction remains constant as the cloud expands, because changing the molecular population requires three-body collisions, which occur with negligible probability during the expansion time. We take the  mean free path to be $\lambda_{mfp}=1/(n_\uparrow\sigma)$, where $n_\uparrow=n/2$ is the central density in one spin state, for a 50-50 mixture. To be conservative, we take $\sigma$ to be the average {\it transport} cross section, $\sigma_{trans}$, with suppressed forward scattering, as used to estimate the viscosity in Ref.~\cite{BruunViscosity2}. For s-wave scattering, we have $\sigma(k) =4\pi a^2/(1+k^2a^2)$. Using a Maxwell-Boltzmann distribution of relative wave vectors $k$, the average transport cross section is then $\bar{\sigma}_{trans}=(2\lambda_T^2/3)\,F(q)$, where $\lambda_T=h/\sqrt{2\pi mk_BT}$ is the thermal wavelength.  $F(q)\equiv \frac{1}{2}\int_0^\infty \frac{dy\,y^3\,e^{-y}}{y+q^2}$, with $q=\lambda_T/(|a|\sqrt{2\pi})$.  For $\lambda_T>>|a|$, $\bar{\sigma}_{trans}\rightarrow 4\pi a^2$. Note that at resonance, $\sigma_{trans}$ is a factor of $6$ smaller than than the thermal average unitary collision cross section and therefore increases the Knudsen number by a factor of $6$  compared to an estimate based on the collisional mean free path~\cite{NJPReview}. Using $m\omega_x^2 R_x^2=2k_BT$, $E=3k_BT$, and $E_F=(3N\omega_x\omega_y\omega_z)^{1/3}\equiv k_B T_{FI}$ (Fermi energy of an ideal gas at the trap center), we find the initial Knudsen number at the cloud center, just after release
\begin{equation}
K_n=\frac{\sqrt{\pi}}{2(3N\lambda_x)^{1/3}F(q)}\left(\frac{E}{E_F}\right)^2.
\label{eq:Knudsen}
\end{equation}
Here, $q\equiv \frac{1}{k_{FI}|a|}\sqrt{\frac{6 E_F}{E}}$ and $\lambda_x\equiv \omega_y\omega_z/\omega_x^2$.

\begin{figure}
\begin{center}\
\includegraphics[width=4.0in]{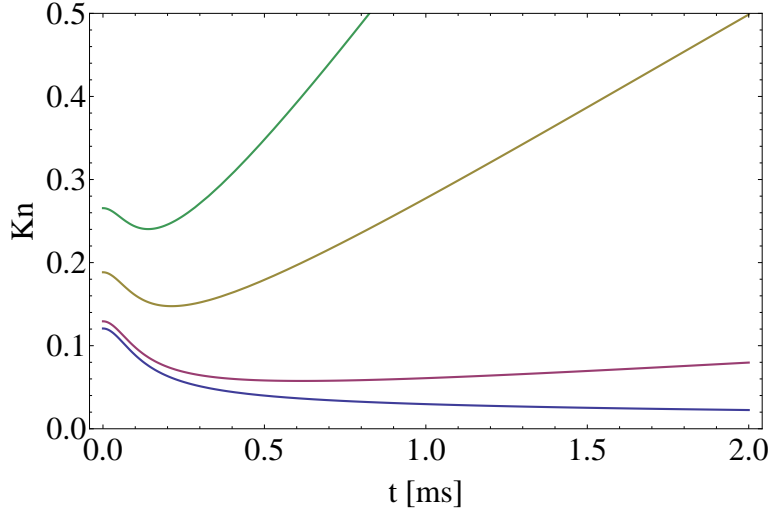}
\end{center}
\caption{Knudsen number at the cloud center as a function of expansion time for different scattering lengths $a$ at an energy $\widetilde{E}/E_F=1.6$. Curves from top to bottom for $1/(k_{FI}|a|)=0.9,0.6,0.2,0$, respectively. \label{fig:Knudsen}}
\end{figure}

As the gas expands, several factors causes the Knudsen number $K_n=\lambda_{mfp}/(2R_x)$ to change. First, the cloud radius $R_x$ increases as $b_x(t)$, with $b_x$ a time-dependent scale factor, as described in detail in \S~\ref{sec:scaling}. Second, the mean free path $\lambda_{mfp}=2/(n\sigma)$ changes, since the density $n$ decreases as $1/\Gamma(t)$, where $\Gamma$ is the volume scale factor. Using an adiabatic approximation, the temperature decreases as $1/\Gamma^{2/3}(t)$, which causes the cross section $\sigma$ to change as well. For the resonantly interacting gas, where $q=0$ and $F(0)=1$, the net effect is that the Knudsen number {\it decreases} as the gas expands, as shown in Fig.~\ref{fig:Knudsen} for $1/(k_{FI}|a|)=0$ by the lowest (blue) curve. This is easy to understand: At resonance, the cross section increases as $\lambda_T^2\propto\Gamma^{2/3}$, so that the Knudsen number then decreases as $K_n(t)=K_n(0)\Gamma^{1/3}(t)/b_x(t)$. Hence, the resonantly interacting gas becomes {\it more} hydrodynamic as the gas expands.  For finite scattering length, the Knudsen number decreases with expansion time until $\lambda_T>|a|$, when the cross section becomes constant, $\rightarrow 4\pi a^2$, and then increases as the density decreases.

\begin{figure}
\begin{center}\
\includegraphics[width=4.0in]{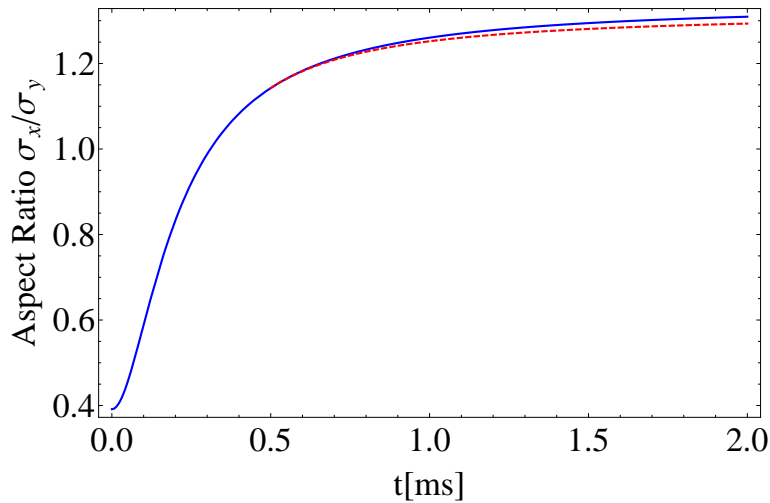}
\end{center}
\caption{Breakdown of hydrodynamic expansion: Aspect ratio versus time for $\widetilde{E}/E_F=1.6$ where the viscosity coefficient $\alpha_{S0}\simeq 3.0$.  Blue solid curve: Hydrodynamic theory; Red dashed curve: Ballistic expansion with initial conditions set by the hydrodynamic theory at $0.5$ ms. \label{fig:HydroToBallistic}}
\end{figure}

We see from Fig.~\ref{fig:Knudsen} that for times $t>0.5$ ms, only the largest $1/(k_{FI}|a|)=0.9$ may deviate from hydrodynamic flow at the highest energies. To investigate the time-dependent breakdown of hydrodynamics, we simulate hydrodynamic flow and abruptly switch to ballistic flow for $t\geq 0.5$ ms and compare the result to hydrodynamic flow for the entire time period.   Fig.~\ref{fig:HydroToBallistic} shows that there is very little difference between switching to ballistic flow for $t\geq 0.5$ ms (red dashed curve) and hydrodynamic flow at all times (blue solid curve). Hence, the asymptotic  aspect ratio is  determined by the hydrodynamic expansion at short times.  If the expansion deviated from hydrodynamic flow by becoming ballistic, we would expect an apparent increase in the shear viscosity, in contrast to the suppression that is observed near $1/(k_{FI}|a|)=0.9$. Further, for a small  dimer fraction, we would expect the Knudsen number to be smaller than that of a cloud comprised solely of atoms.  We do not observe any abrupt changes in the aspect ratio versus time data for the normal fluid regime studied in the expansion experiments. Hence, we assume that a  hydrodynamic description is satisfactory and that dimer-atom mixtures expand as a single fluid.

\subsubsection{Effects of Bulk Viscosity $\langle \alpha_B \rangle$ and $\Delta p$ on the Aspect Ratio}
\label{sec:highT}

The bulk viscosity and the conformal symmetry breaking pressure affect the expansion of the gas in a similar way,
as both involve scalar quantities. In Ref.~\cite{ElliottScaleInv}, we measured the bulk viscosity to be $\langle \alpha_B \rangle \leq 0.04\,\hbar n$ on resonance.  Further, we found that the difference between the off-resonance and on-resonance expansion dynamics is dominated by the conformal symmetry breaking pressure $\Delta p$. The effect of the off-resonance bulk viscosity is much smaller.  For this reason, we focus in this section on the $\Delta p$ correction.

When the bias magnetic field is tuned away from the Feshbach resonance, the pressure deviates from the unitary limit  $\Delta p=p-\frac{2}{3}{\cal E}$. We show that $\Delta p$ has a negligible effect on the transverse aspect ratio, $\sigma_x/\sigma_y$, compared to the shear viscosity. This is accomplished using a simple model. Based on dimensional analysis, to first order in in $1/(k_Fa)$, $\Delta p$ has a natural scale $n\epsilon_F(n)/(k_Fa)$, where $\epsilon_F(n)\propto k_F^2$ is the local Fermi energy.  Hence, $\Delta p\propto k_F$  requires the time dependence $\Gamma^{-1/3}$, so we take
\begin{equation}
\frac{1}{N}\int d^3{\mathbf{r}}\,\Delta p=C\,\frac{\langle{{\mathbf{r}}}\cdot\nabla U_{total}\rangle_0}{3k_{FI}a}\,\Gamma^{-1/3}(t),
\label{eq:Deltap}
\end{equation}
where $C$ is a constant.  As shown in Ref.~\cite{ElliottScaleInv}, the next order (quadratic) term in $1/(k_Fa)$ is time independent, and has no effect on the expansion dynamics. Using Eq.~\ref{eq:Deltap} in Eqs.~\ref{eq:Cdeltap}-\ref{eq:6.5e}, we then obtain
\begin{equation}
C_{\Delta p}(t)=-\frac{C}{k_{FI}a}\,[\Gamma^{1/3}(t)-1]
\label{eq:C}
\end{equation}
We determine $C=0.07$ from the measured expansion of the mean square cloud radius $\langle{\mathbf{r}}^2\rangle=\langle x^2+y^2+z^2\rangle$ both on and off resonance, with $1/(k_{FI}a)=0,\pm 0.6$ and $\tilde{E}/E_F =1.0$, as shown in Fig.~4 of Ref.~\cite{ElliottScaleInv}. We then use Eq.~\ref{eq:C} in Eq.~\ref{eq:6.1e} to determine the cloud radii and the transverse aspect ratio as a function of time after release.

\begin{figure}[htb]
\begin{center}\
\hspace*{-0.25in}\includegraphics[width=5.0in]{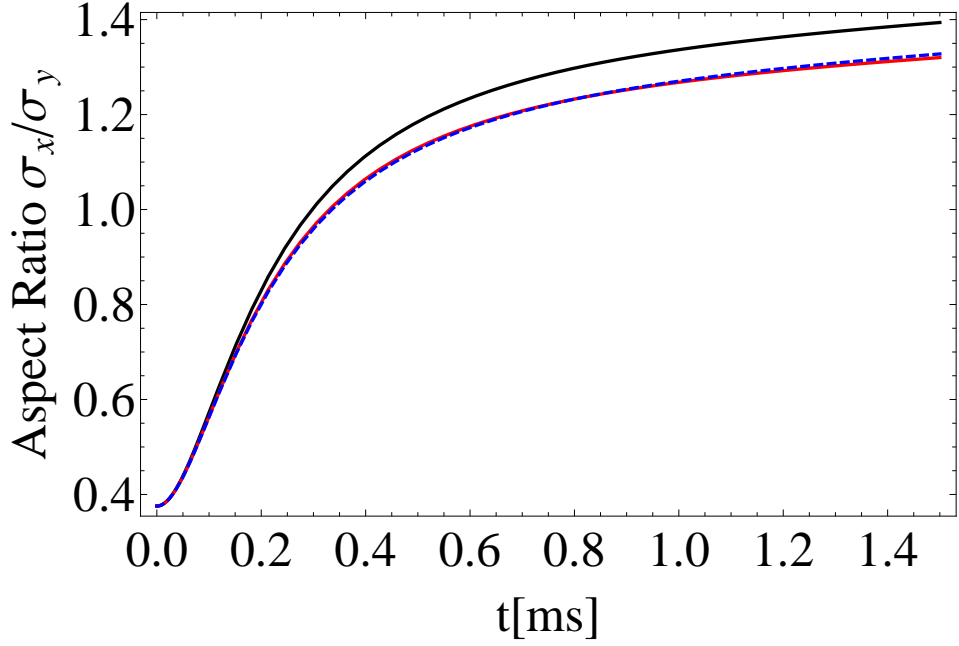}
\end{center}
\caption{Transverse aspect ratio $\sigma_x/\sigma_y$ for $1/(k_{FI}a)=+0.6$ as a function of time after release, calculated using Eq.~\ref{eq:6.1e} for $\widetilde{E}/E_F=1.0$, where the measured viscosity at resonance is $\langle \alpha_S \rangle_0=1.0$. Black solid line unitary result: The change in the shear viscosity relative to the unitary value $\Delta \langle \alpha_S \rangle=0$, $C_{\Delta p}=0$; Red solid line, off-resonance for $\Delta p=0$: $\Delta \langle \alpha_S \rangle=0.5$, $C_{\Delta p}=0$. Blue dashed line, off-resonance result with $\Delta p\neq 0$: $\Delta\langle \alpha_S \rangle = 0.5$, $C_{\Delta p}$ determined from Eq.~\ref{eq:C} with $C=0.07$ (as measured from the expansion of the corresponding mean square cloud radius),  showing negligible effect on $\sigma_x/\sigma_y$.\label{fig:A1a}}
\end{figure}

Fig.~\ref{fig:A1a} shows that the effect of $\Delta p$ on the $\sigma_x/\sigma_y$ aspect ratio. As in the main text, we define $\Delta \langle \alpha_S \rangle$ as the change in the shear viscosity relative to the unitary value $\langle\alpha_S\rangle_0$ at the same $\widetilde{E}$. We see that the effect of $\Delta p$  is negligible compared to the effect from the change in the shear viscosity. For this reason, we neglect both $\Delta p$ and the bulk viscosity in our determination of the shear viscosity in the off-resonance regime.

\subsection{Measurement of the Shear Viscosity}
In the experiments the cloud radii are measured as a function of time after release in all three dimensions, using two simultaneous probe pulses interacting with different spin states to obtain independent absorption images on two CCD cameras~\cite{ElliottScaleInv}.  The parameters measured in our experiments are the cloud radii $\sigma_i=\sqrt{2\langle x_i^2 \rangle}$ and number of atoms $N$. Using Eq.~\ref{eq:4.6e} in Eq.~\ref{eq:6.1e}, and ignoring the effects of the conformal symmetry breaking pressure and bulk viscosity as discussed above, the equations of motion for the scale factors  reduce to
\begin{equation}
\ddot{b}_i=\frac{\langle x_i\partial_i U\rangle_0}{m\langle x_i^2\rangle_0\Gamma^{2/3}b_i\,}\left[1+ C_Q(t)\right]
-\frac{\hbar\langle \alpha_S \rangle\sigma_{ii}}{m\langle x_i^2\rangle_0 b_i},
-\omega^2_{i\,mag}b_i,
\label{eq:ShearEq1}
\end{equation}
where $\Gamma\equiv b_xb_yb_z$. The last term arises from the bias magnetic field curvature, where $\omega^2_{y\,mag}=\omega^2_{z\,mag}\equiv\omega^2_{mag}$ and $\omega^2_{x\,mag}=-2\,\omega^2_{mag}$, with  $\omega^2_{mag}$ given in the main text. The heating term is determined from
\begin{equation}
\dot{C}_Q\equiv\frac{\hbar\,\langle \alpha_S \rangle\Gamma^{2/3}}{\langle {\mathbf{r}}\cdot\nabla U\rangle_0}\sum_i\sigma_{ii}^2,
\label{eq:ShearEq2}
\end{equation}
 with
\begin{eqnarray*}
\sigma_{ii} &=&2\frac{\dot{b}_i}{b_i}-\frac{2}{3}\frac{\dot{\Gamma}}{\Gamma}\\
\sum_i\sigma_{ii}^2&=&4\sum_i\frac{\dot{b}_i^2}{b_i^2}-\frac{4}{3}\frac{\dot{\Gamma}^2}{\Gamma^2}.
\end{eqnarray*}
Eq.~\ref{eq:ShearEq1} and Eq.~\ref{eq:ShearEq2} constitute a set of four differential equations that can be solved numerically with initial values $b_i(0) = 1$, $\dot{b}_i(0) = 0$, and $C_Q(0) = 0$, providing the expansion factors $b_i$ as a function of time after release of the cloud.

In the experiments, the shear viscosity is parameterized by the scattering length-independent energy scale $\widetilde{E}\equiv\langle{\mathbf{r}}\cdot\nabla U\rangle_0$, as described in the main text.  For a given value of $\langle \alpha_S \rangle$, we can then determine $\langle x^2_i \rangle_0 = \langle x^2_i\rangle / b_i^2$ as well as $\langle x_i\partial_i U_{total}\rangle_0$ and $\langle \mathbf{r}\cdot\nabla U\rangle_0$. Consistency is tested by checking that $\langle x_i\partial_iU\rangle_0$ is the same for all directions $i=x,y,z$, which follows from force balance in the trap for a scalar pressure.  In the harmonic approximation this requires
\begin{equation*}
\omega_x^2 \langle x^2 \rangle_0 = \omega_y^2 \langle y^2 \rangle_0.
\end{equation*}
With an anharmonic trap,  we require instead
\begin{equation*}
\omega_x^2 \langle x^2 \rangle_0[1+f_x(\widetilde{E})] = \omega_y^2 \langle y^2 \rangle_0[1+f_y(\widetilde{E})],
\end{equation*}
where the $f_{x,y}(\widetilde{E})$ are anharmonic correction factors. For the transverse directions $x$ and $y$, we assume identical correction factors, $f_x(\widetilde{E})=f_y(\widetilde{E})$. Then, for both harmonic and anharmonic traps, the transverse aspect ratio as a function of time is determined by
\begin{equation}
\sqrt{\frac{\langle x^2 \rangle}{\langle y^2 \rangle}} = \sqrt{\frac{\langle x^2 \rangle_0 b_x^2}{\langle y^2 \rangle_0 b_y^2}} = \frac{b_x}{b_y} \frac{\omega_y}{\omega_x}.
\end{equation}

\begin{figure}[htb]
\begin{center}\
\includegraphics[width=5in]{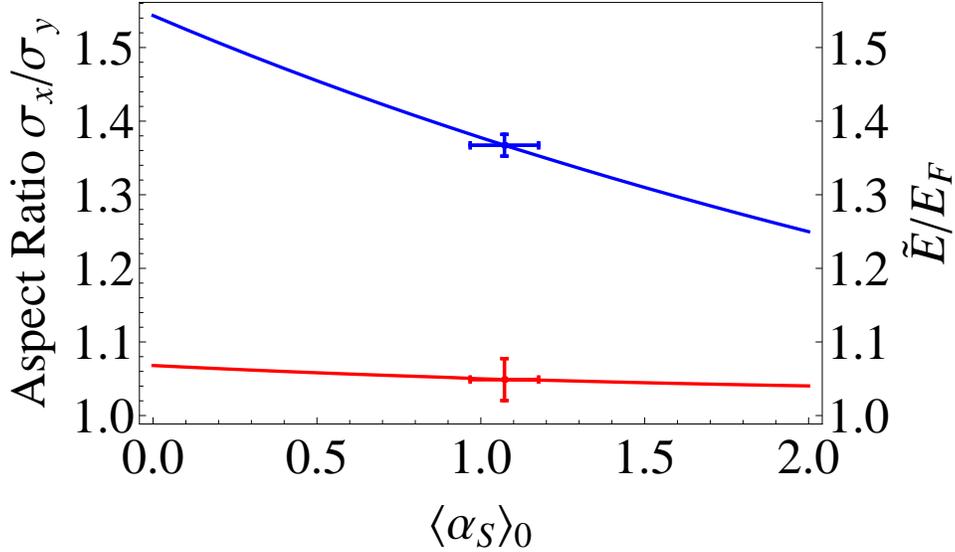}
\end{center}
\caption{Variation of the self-consistently computed energy and aspect ratio $\sigma_x/\sigma_y$  with the trial shear viscosity coefficient $\langle\alpha_S \rangle_0$. The aspect ratio (blue curve) is determined for a fixed time of $1.2$ ms after release. The energy $\widetilde{E}$ (red curve) is found from the initial cloud radii, which  are determined from the  cloud radii measured at 1.2 ms using the calculated expansion factors, which vary with the shear viscosity.   Note that the aspect ratio is very sensitive to the shear viscosity, while the energy is not.  The dots with error bars are the  corresponding self-consistent values for a single measurement of the three cloud radii.
\label{fig:S1}}
\end{figure}

Eq.~\ref{eq:ShearEq1} and Eq.~\ref{eq:ShearEq2} provide expansion factors $b_i$ that depend upon $\langle \alpha_S \rangle$ and $\langle x^2_i\rangle_0$, which in turn depends upon the expansion factors. These parameters are determined self consistently by iterative fits to the aspect ratio data, using initial guesses and then reiterating until the desired precision is achieved. Fig.\ref{fig:S1} shows curves corresponding to the model aspect ratio $\left(b_x/b_y\right)/\left(\omega_x/\omega_y\right)$ at $1.2$ $ms$ after the trap has been turned off and the corresponding measured $\widetilde{E}$ as a function of a test $\langle \alpha_S \rangle$ for a single data point.  Note the sensitivity of aspect ratio to shear viscosity and the relative insensitive of energy to shear viscosity.

As described in the main text, the shear viscosity coefficient varies with time as
\begin{equation}
\langle \alpha_S \rangle = \langle \alpha_S \rangle_0 +\frac{c_1}{k_{FI} a}\, \Gamma ^{1/3}+\frac{c_2}{\left(k_{FI} a \right)^2}\, \Gamma^{2/3}
\label{eq:ShearFull}
\end{equation}
To determine the coefficients $c_1$ and $c_2$ from the $\widetilde{E}$ and $1/k{FI}$ dependent aspect ratio data,  we fit the data globally.

\begin{figure}[htb]
\begin{center}\
\includegraphics[width=5in]{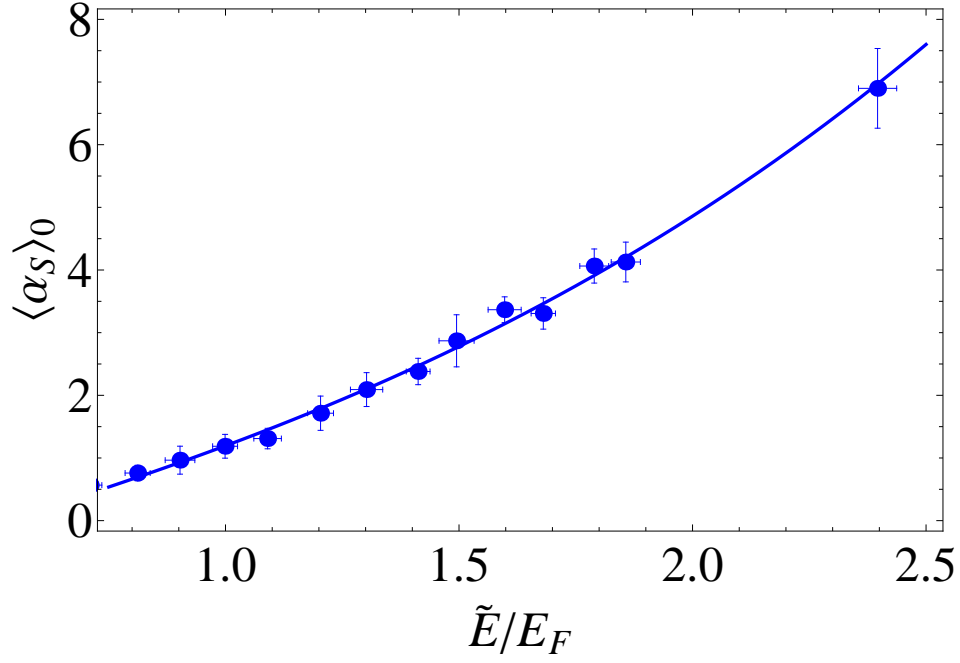}
\end{center}
\caption{Unitary shear viscosity data (points) $\langle \alpha_S \rangle_0$ as a function of $\widetilde{E}/E_F$.  301 data points are The data is divided into energy bins with $0.1 \widetilde{E}/E_F$ spacing and averaged.  Blue line is a polynomial $d_0 + d_1 \widetilde{E}/E_F + d_2 (\widetilde{E}/E_F)^2$ where $d_0 = -0.31$, $d_1 = 0.35$, and $d_2 = 1.14$ provides the best fit to the data.
\label{fig:S3}}
\end{figure}

First, we find the shear viscosity coefficient at resonance $\langle \alpha_S \rangle_0$ as a smooth function of energy $\widetilde{E}/E_F$. To measure the energy dependence, 301 data points are divided into energy bins with $0.1 \widetilde{E}/E_F$ spacing and averaged. Then the resulting averaged data, Fig.~\ref{fig:S3}, are fit with the polynomial $d_0 + d_1 \widetilde{E}/E_F + d_2 (\widetilde{E}/E_F)^2$, where the best fit gives $d_0 = -0.31$, $d_1 = 0.35$, and $d_2 = 1.14$, which is shown as the smooth curve.
Next, we use Eq.~\ref{eq:ShearFull} in Eq.~\ref{eq:ShearEq1} and Eq.~\ref{eq:ShearEq2} to find $c_1$ and $c_2$, where the polynomial fit for the  viscosity at resonance determines $\langle \alpha_S \rangle_0$ as a function of $\widetilde{E}/E_F$. $c_1$ and $c_2$ are determined using a $\chi ^2$ fit to the aspect ratio data.  As noted in the text,  off-resonance, the change in the shear viscosity  depends on both $\widetilde{E}$ and $1/(k_{FI} a)$ and the data deviates from the shifted parabolic fit at extreme values of $1/(k_{FI} a)$.  Therefore we divide the data into discrete energy ranges as in the resonant case and also limit the range of interaction strength to $-0.5 < 1/(k_{FI} a) < 0.7$ to avoid the extreme values of $1/(k_{FI} a)$.  The  results of the 2-parameter $\chi^2$ fit for $c_1$ and $c_2$ are summarized in Table~\ref{tab:T1}.  Errors $\Delta c_1$ (holding $c_2$ constant) and $\Delta c_2$ (holding $c_1$ constant) are estimated from the range where the normalized $\chi^2$ increases by the inverse of the number of data points (\# pts) in the fit.  $\Delta \widetilde{E}/E_F$ is simply the standard deviation for the measured energy range.

\begin{table}[htb]
\large
\begin{tabular}{|c|c|c|c|c|c|c|c|c|c|}
\hline
Energy Range & \# pts & $\widetilde{E}/E_F$ & $\Delta\widetilde{E}/E_F$ & $\chi^2$ & $c_1$ & $\Delta c_1$ & $c_2$ & $\Delta c_2$ & $\frac{-c_1}{2 c_2}$ \\ \hline
0.75-0.90  & 78    & 0.83                & 0.05                       & 2.2                   & -1.43 & 0.07         & 2.49  & 0.10         & 0.29                  \\ \hline
0.90-1.10    & 71    & 0.97                & 0.06                       & 1.1                   & -1.22 & 0.05         & 2.42  & 0.09         & 0.25                  \\ \hline
1.10-1.30    & 123   & 1.18                & 0.05                       & 2.0                   & -1.11 & 0.07         & 2.19  & 0.09         & 0.25                  \\ \hline
1.30-1.65    & 57    & 1.40                & 0.08                       & 1.4                   & -0.87 & 0.12         & 2.05  & 0.18         & 0.21                  \\ \hline
1.65-2.50    & 19    & 1.90                 & 0.25                       & 1.4                   & -0.36 & 0.12          & 1.52  & 0.22         & 0.11                  \\ \hline
\end{tabular}
\label{tab:T1}
\caption{Determination of $c_1$ and $c_2$ in Eq.~\ref{eq:ShearFull}. The first column shows the range of each energy bin. $\chi^2$ is the total $\chi^2$ normalized by the number of data points in each bin. The last column gives the location of the center of the parabolic fit versus $1/(k_{FI}a)$ for each energy bin.}
\end{table}


\begin{thebibliography}{33}
\bibitem[{Res()}]{ResonantE}
\bibinfo{note}{For the resonantly interacting gas, where $p = 2{\cal E}/3$ with
  ${\cal E}$ the local energy density~\cite{HoUniversalThermo},
  $\widetilde{E}/2$ is just the internal energy per
  particle~\cite{ThomasUniversal}. At resonance, the total energy is then
  $E=\langle U\rangle_0+\widetilde{E}/2$, with $\langle
  U\rangle_0\equiv\frac{1}{N}\int d^3{\mathbf{r}} \,n_0({\mathbf{r}})\,
  U({\mathbf{r}})$~\cite{ThermoLuo}.}
\bibitem[{Neg()}]{NeglectBulk}
\bibinfo{note}{We ignore the bulk viscosity, which is found to be much smaller
  than shear viscosity~\cite{ElliottScaleInv,SupportOnline}.}
\end{thebibliography}
\end{document}